\documentclass[twocolumn,showpacs,superscriptaddress]{revtex4}

\usepackage{graphicx}%
\usepackage{dcolumn}
\usepackage{amsmath}
\usepackage{amssymb}

\begin{document}

\newcommand{\para}{\ensuremath{/\!/}}
\newcommand{\MgB}{MgB$_2$}
\newcommand{\Hcab}{$H_{c2}^{(ab)}$}
\newcommand{\Hcc}{$H_{c2}^{(c)}$}
\newcommand{\Hc}{$H_{c2}$}
\newcommand{\etal}{{\it et al.}}
\newcommand{\Nature}[3]{Nature {\bf {#1}}, {#2} ({#3})}
\newcommand{\PRL}[3]{Phys. Rev. Lett. {\bf {#1}}, {#2} ({#3})}
\newcommand{\PRB}[3]{Phys. Rev. B {\bf {#1}}, {#2} ({#3})}
\newcommand{\PC}[3]{Physica C {\bf {#1}}, {#2} ({#3})}
\newcommand{\EPL}[3]{Europhys. Lett. {\bf {#1}}, {#2} ({#3})}

\preprint{Version 5.02} 

\title[Specific heat of single crystal MgB$_2$]{Specific heat of single crystal MgB$_2$: \\
a two-band superconductor with two different anisotropies}

\author{F. Bouquet}
\author{Y. Wang}
\author{I.~Sheikin}
\author{T.~Plackowski}
\author{A.~Junod}

\affiliation{%
DPMC, University of Geneva, 24 quai Ernest-Ansermet, 1211 Gen\`eve
4, Switzerland
}%

\author{S.~Lee}
\author{S.~Tajima}

\affiliation{%
Superconductivity Research Laboratory, ISTEC, 1-10-13 Shinonome,
Koto-ku, Tokyo 135-0062, Japan
}%

\date{\today}

\begin{abstract}
Heat-capacity measurements of a 39~$\mu$g MgB$_2$ single crystal
in fields up to 14~T and below 3~K allow the determination of the
low-temperature linear term of the specific heat, its field
dependence and its anisotropy. Our results are compatible with
two-band superconductivity, the band carrying the smaller gap
being isotropic, that carrying the larger gap having an anisotropy
of $\sim 5$. Three different upper critical fields are thus needed
to describe the superconducting state of MgB$_2$.
\end{abstract}


\maketitle

Shortly after the discovery of 40-K superconductivity in
MgB$_2$~\cite{Nagamatsu}, its nature was intensively studied. The
isotope effect  soon indicated phonon-mediated
pairing~\cite{Canfield}. However, MgB$_2$ cannot be considered as
a classic conventional superconductor; indeed several studies
point to the existence of a gap much smaller than the expected BCS
value: band-structure calculations~\cite{Shulga,Liu},
STM~\cite{Hermann,Morten}, specific heat
($C$)~\cite{Wang,BouquetPRL,JunodNarlikar,Europhys,Yang},
penetration depth~\cite{Carrington}, and various spectroscopic
experiments~\cite{Chen,Quilty,Tsuda,Giubileo,Szabo,Laube}. The
so-called ``two-band'' model~\cite{Shulga,Liu}, which considers
two different $s$-wave superconducting gaps (a larger and a
smaller gap, $\Delta_{\sigma}$ and $\Delta_{\pi}$) on separate
sheets of the Fermi surface (the $\sigma$- and $\pi$-band,
respectively), now seems widely accepted and explains most
experimental results~\cite{Golubov,Choi}, with few
exceptions~\cite{NMR}. The theory of two-band superconductivity
was suggested soon after the development of the BCS
theory~\cite{Suhl}, and some s-d metals have shown hints of this
phenomenon, generally in the form of small deviations from BCS
predictions, e.g. in $C$ experiments~\cite{Phillips}. In contrast,
MgB$_2$ shows such clear signatures that it might serve as a
textbook example. For instance, the electronic specific heat of
MgB$_2$ shows an excess by an order of magnitude  at
$\sim$~$T_c/5$, and presents an exponential dependence at low
temperature ($T$) indicating the existence of a small gap on a
large fraction ($\sim$~50\%) of the Fermi
surface~\cite{Wang,BouquetPRL,JunodNarlikar,Europhys,Yang}. At low
$T$ the effect of a small magnetic field ($H$) on the coefficient
of the linear $C$-term,  $\gamma(H) \equiv
\lim_{T\rightarrow0}C(H)/T$, is dramatic. Classically,
$\gamma(H)$, a thermodynamic probe of the  electronic density of
states at low-energy, is expected to be linear in $H$ up to
\Hc~\cite{Caroli}. However, for MgB$_2$ an extreme non-linearity
was observed, which was assigned to the existence of an
additional, smaller gap~\cite{Wang,BouquetPRL,JunodNarlikar,Yang}.

The only specific heat experiment on single crystal reported so
far was focused on the superconducting
transition~\cite{Welp,Lyard}. Such experiments are delicate since
the mass of MgB$_2$ crystals is usually smaller than 100~$\mu$g,
but are required to study the anisotropy. Previous studies showed
that the anisotropy of superconducting properties depends on both
field and temperature~\cite{Angst,Eltsev}. This puzzling behavior
motivated the present study.

We measured the low-temperature specific heat of a MgB$_2$ single
crystal with the field parallel and perpendicular to the boron
planes. The results clearly show that the extreme non-linearity of
$\gamma(H)$, first observed in
polycrystals~\cite{Wang,BouquetPRL,JunodNarlikar,Yang}, is an
intrinsic property of MgB$_2$, related to the existence of two
superconducting gaps. Moreover, these measurements reveal a
dramatic variation of the effective anisotropy with the field,
from 1 to $\sim$~5. We interpret these results as a manifestation
of the two-band nature of  superconductivity in MgB$_2$, plus the
dimensionality of the $\pi$-band (isotropic) and $\sigma$-band
(anisotropic).


\begin{figure}
\includegraphics{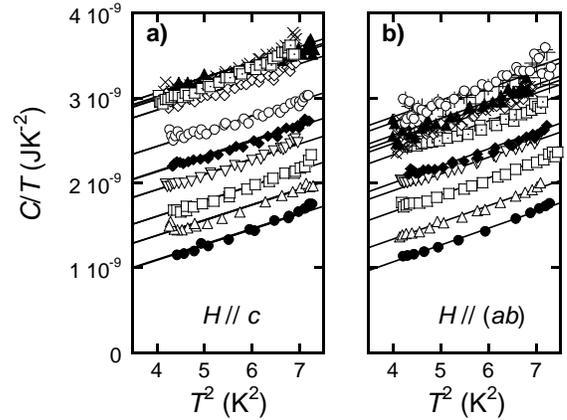}
\caption{Typical curves of total specific heat, including that of
addenda, for different magnetic fields (from bottom to top); {\bf
a)} 0, 0.1, 0.2, 0.5, 1, 2, 3, 4, 6, and 10~T  along the $c$-axis;
{\bf b)} 0, 0.1, 0.2, 0.5, 1, 3, 6, 8, 10, 12, and 14~T in the
$(ab)$-planes. The straight lines are linear fits with a common
slope.}
\label{FigCsurT}%
\end{figure}

The single crystal of  MgB$_2$ is similar to those described in
Ref.~\cite{Tajima}. Its mass, 39~$\mu$g, is   small for a
quantitative study of specific heat. For this purpose, we
developed  a miniaturized version of the relaxation calorimeter
described in Ref.~\cite{Wang}. A commercial Cernox bare
chip~\cite{LakeShore} was thinned down to 1~mg mass and used
simultaneously as  sample platform,  heater and  thermometer. Two
10~$\mu$m phosphor-bronze wires soldered onto the chip acted as
mechanical support, thermal links and electrical leads. Only
low-$T$ studies were possible, since the heat capacity of the
platform, essentially of phononic origin (sapphire),  increases
rapidly with $T$. We thus restrained our study to the range 2 to
3~K. Previous experiments have shown that this range of
temperature lies sufficiently above the domain of possible
parasitic magnetic contributions, but low enough  ($\sim T_c/15$)
to allow reliable determination of $\gamma(H)$ (see e.g.
Ref.~\cite{JunodNarlikar}). Owing to the small value of $C$ at
these temperatures, the relaxation time was short (0.1--0.5~s),
necessitating aperture times shorter than one power line cycle for
the digitization of voltages. In order to suppress the 50~Hz
noise, the signal was averaged over 50 to 200 relaxations. The
field dependence of the addenda represents at most 13\% of the
normal electronic linear term of the MgB$_2$ crystal, and was
calibrated by measuring a 196~$\mu$g  Ag reference sample.

\begin{figure}
\includegraphics{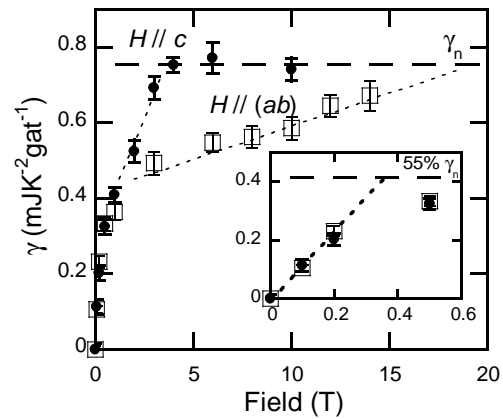}%
\caption{Coefficient of the electronic linear term versus magnetic
field applied parallel ($\square$) or perpendicular ($\bullet$) to
the boron planes. The long-dashed line represents the normal state
contribution. The short-dashed lines are guides for the eyes.
Inset: expanded view of the low-field region; here the long-dashed
line represents the partial normal-state contribution of the
small-gapped band (see text and Fig.~\ref{FigDecomposition}). }
\label{Figgamma}%
\end{figure}

Figure~\ref{FigCsurT} shows the specific heat for both
orientations. It is impossible to compensate exactly for the small
but unknown quantity of GE-varnish used to attach the sample.
Instead, in order to calculate $\gamma(H)$, we used the zero-field
curve as a reference. It was shown on polycrystals that the
electronic specific heat of MgB$_2$ is negligible in zero field in
this temperature range~\cite{BouquetPRL,Yang,JunodNarlikar}: the
parallel shift of the $C/T$ curves in Fig.~\ref{FigCsurT} is thus
a direct measurement of $\gamma(H)$. This shift was precisly
determined by fitting each curve with  a linear electronic term
$\gamma(H)T$ plus a $\beta T^3$ lattice term, imposing the same
$\beta$ coefficient for all fields. The result for $\gamma(H)$ is
plotted in Fig.~\ref{Figgamma} for both orientations. With the
field perpendicular to the boron planes, $\gamma(H)$ saturates
above 3--4~T, indicating that the sample has entered the normal
state. This value is consistent with previous reports giving
$\mu_0H_{c2}^{(c)} \approx$~3~T for the upper critical field along
the $c$-axis ~\cite{Welp,Lyard,Angst,Sologubenko}. This also
determines the normal electronic linear term $\gamma_n = 0.76 \pm
0.03$~mJK$^{-2}$gat$^{-1}$, comparable with literature
values~\cite{Wang,BouquetPRL,Yang,JunodNarlikar}. The normal state
is not fully established when the field is parallel to the
$(ab)$-planes, indicating that $H_{c2}^{(ab)}$ is higher than the
maximum applied field of 14~T. An extrapolation suggests
$\mu_0H_{c2}^{(ab)} \approx$~18--22~T, in agreement with reported
values~\cite{Buzea}. Therefore, the anisotropy of the upper
critical field at $T \ll T_c$ is established in bulk.


The anisotropy of \Hc{} in MgB$_2$ has long been
debated~\cite{Buzea}, but was finally assessed by experiments
using clean single
crystals~\cite{Angst,Eltsev,Welp,Lyard,Sologubenko}. Two kinds of
anisotropy can be considered: the anisotropy of the
superconducting gap, and that of the Fermi surface. To model the
former, some $\vec{k}$-dependence of the gap has to be assumed.
Such an approach was attempted using a one-band model for
MgB$_2$~\cite{Maki}. The two-band model is an extreme case of such
an anisotropy: depending on which Fermi sheet is considered, the
superconducting gap may be either large or small. The effect of an
anisotropic Fermi surface on superconducting properties  can be
modeled by renormalizing the isotropic results by a function of
the ratio of the effective masses, a method that was successfully
applied to cuprate superconductors~\cite{Blatter}. Angst {\it et
al.}{} used this approach for MgB$_2$~\cite{Angst}, but showed
that the shape of the torque curves did not follow this simple
one-band, anisotropic model over the whole range of field
orientations.

\begin{figure}
\includegraphics{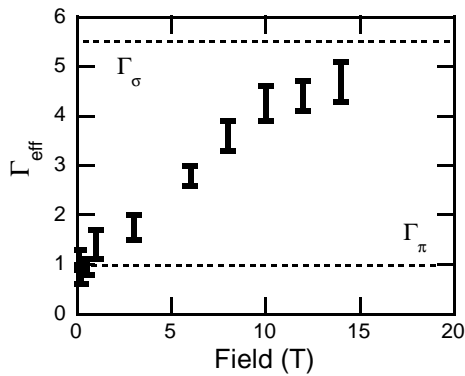}
\caption{Effective anisotropy, defined as the ratio of magnetic
fields applied in the $(ab)$-planes and along the $c$-axis
yielding the same $\gamma$-value. Note that the choice of the
abscissa is arbitrary: we chose $H\para (ab)$, but we could just
as well plot it versus
$H\para c$ or versus $\gamma(H)$.}%
\label{FigGeff}
\end{figure}

From Fig.~\ref{Figgamma}, it is clear that it is not possible to
normalize the effect of the field by using a constant anisotropy
factor. For $H \leq 0.5$~T, the values of $\gamma(H)$ are almost
undistinguishable, irrespective of the field direction, showing
the absence of anisotropy (see inset of Fig.~\ref{Figgamma}).
However, a field scaling factor of $\sim 5$ is needed to superpose
the $\gamma(H)$ curves just below saturation, i.e. near \Hc. An
effective anisotropy $\Gamma_{eff}$ can be defined as the scaling
factor by which the field along the $c$-axis must be multiplied in
order to merge both $\gamma(H)$ curves. Figure~\ref{FigGeff}
presents $\Gamma_{eff}$ versus $H\para(ab)$. At low field,
$\Gamma_{eff} \approx$~1, since the effect of $H$ does not
measurably depend on the orientation; $\Gamma_{eff}$ rapidly
increases with   $H$, and tends toward the anisotropy of \Hc,
$\sim 5$. Such an $H$-dependent anisotropy was also deduced from
torque measurements~\cite{Angst}. The present analysis shows that
the change of $\Gamma_{eff}$ with $H$ (and possibly with $T$) is
an intrinsic property of MgB$_2$, having a bulk thermodynamic
signature, even though the amplitude of this variation may  depend
on the physical quantity by which it is determined. We will argue
in the following that, to explain specific heat measurements, both
kinds of anisotropy are required: the band dependence of the gap,
{\it plus} a renormalization by the effective masses on the
$\sigma$-band.


Let us first focus on the shape of the $\gamma(H)$ curves.
Similarly to previous data on polycrystals, $\gamma(H)$ increases
sharply at low fields. The present data show that this effect does
not depend on the orientation of the field, and point to the
existence of a large fraction of normal electrons in the mixed
state already at very low $H$. In a typical $s$-wave, dirty
type-II superconductor, $\gamma(H)$ mainly originates from the
vortex cores and should be approximately proportional to $H$ (i.e.
the number of vortices). The strong non-linearity of $\gamma(H)$
for MgB$_2$ is atypical: for $\mu_0H=1$~T, approximatively half of
$\gamma_n$ is recovered, irrespective of the field direction. The
effect is particularly striking for $H
\para (ab)$: about 5\% of \Hcab drives $\sim$~50\% of the
electrons into the normal state.

The peculiar shape of $\gamma(H)$ can be understood by considering
the meaning of a vortex within a two-band model. Both bands supply
the carriers and currents that define one vortex, and both bands
contribute to the local density-of-states (LDOS) in one vortex
core. Simulations~\cite{Nakai} and STM studies~\cite{Morten}
suggest that a core results from the superposition of two peaks in
the LDOS with different diameters defined as usual by the
coherence length $\xi_{\sigma}$ or $\xi_{\pi}$ associated with
each band. The sharper peaks with diameter $\xi_{\sigma}$ lead to
a conventional core contribution for the $\sigma$-band, and
overlap at \Hc. In contrast, the large coherence length of the
$\pi$-band gives rise to broad LDOS peaks, with a diameter  much
larger than $(\Phi_0/2\pi H_{c2})^\frac{1}{2}$. These giant cores,
which have been observed by STM~\cite{Morten}, contribute heavily
to the total low-$T$ DOS, as measured by $\gamma(H)$, until they
start to overlap much below \Hc. At this point, the contribution
of the $\pi$-band to $\gamma(H)$ saturates. At higher fields,
superconductivity in the $\pi$-band is maintained up to \Hc{} by
coupling to the $\sigma$-band, but the variation of $\gamma(H)$
now comes from the $\sigma$-band contribution. The $\gamma(H)$
curves of Fig.~\ref{Figgamma} compare qualitatively well with
these predictions. Note that the analysis of previous measurements
on polycrystalline
samples~\cite{Wang,BouquetPRL,JunodNarlikar,Yang} was complicated
by angular averaging of $\gamma(H \para c)$ and $\gamma(H\para
ab)$, leading to a flattening above 8~T.


\begin{figure}
\includegraphics{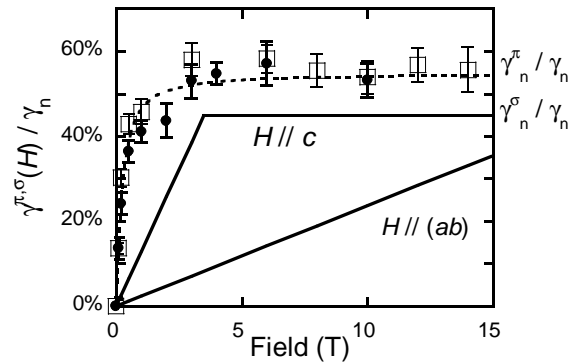}
\caption{Separation of both contributions to $\gamma(H)$:
$\gamma^{\sigma}(H)$ (straight lines) and $\gamma^{\pi}(H)$ for
the field applied along the $c$-axis ($\bullet$) and in the
$(ab)$-planes ($\square$). The dashed line is a guide to the eyes for $\gamma^{\pi}(H)$.}%
\label{FigDecomposition}
\end{figure}

Within the present scheme, the data of Fig.~\ref{Figgamma} can be
used to determine the anisotropy $\Gamma_{\pi}$ and
$\Gamma_{\sigma}$ of each subsystem separately. The abnormally
fast initial increase of $\gamma(H)$ at low $H$ is attributed to
the giant vortex cores associated with $\pi$-band carriers, and,
as shown by the inset of Fig.~\ref{Figgamma}, is isotropic;
therefore $\Gamma_{\pi}=1$. On the other hand, the $\sigma$-band
contribution is responsible for the anisotropy of \Hc, which we
determined when describing Fig. 2; therefore $\Gamma_{\sigma}
\approx 6$. This is consistent with the 3D and quasi-2D character
of the $\pi$- and $\sigma$-band, respectively, found in
band-structure calculations~\cite{Liu,Golubov,Choi}, and with the
experimental determination by torque magnetometry at low
$T$~\cite{Angst}. The {\it effective} anisotropy shown in
Fig.~\ref{FigGeff} is a weighted average which depends on the
relative contribution of each band to the condensate for a given
$H$ and $T$: it varies smoothly from $\Gamma_{\pi}$ at low fields
to $\Gamma_{\sigma}$ at high fields.

Besides the anisotropy of each subsystem, we can attempt to
determine the full shape of the unusual contribution of the
$\pi$-band, $\gamma^{\pi}(H)$. For this purpose, we assume that
the $\sigma$-band contribution, $\gamma^{\sigma}(H)$, follows a
classical behavior~\cite{Caroli}, increasing {\it linearly}  from
0 to $\gamma^{\sigma}_n$ as the field goes from 0 to \Hcc or
\Hcab, depending on the field orientation. The linear sections
above 1~T in Fig.~\ref{Figgamma} correspond to this contribution
of the $\sigma$-band; their extrapolation to $H = 0$ points to the
normal-state value $\gamma^{\pi}_n$ of the $\pi$-band contribution
($\gamma^{\pi}_n+\gamma^{\sigma}_n=\gamma_n$). By subtracting
$\gamma^{\sigma}(H)$ from $\gamma(H)$, we isolate
$\gamma^{\pi}(H)$ (Fig.~\ref{FigDecomposition}). The parameters
\Hcc, \Hcab, $\gamma^{\pi}_n$, and $\gamma^{\sigma}_n$ are
determined from Fig.~\ref{Figgamma}. By requiring that the
saturation value of $\gamma^{\pi}(H)$ is isotropic, we find that
the parameters cannot depart by more than $\pm 15$\% from the
average fitted values $\mu_{0}H_{c2}^{(c)}= 3.5$~T,
$\mu_{0}H_{c2}^{(ab)}= 19$~T, $\gamma^{\pi}_n =
0.42$~mJK$^{-2}$gat$^{-1}$ and $\gamma^{\sigma}_n =
0.34$~mJK$^{-2}$gat$^{-1}$. The $\gamma^{\pi}_n/
\gamma^{\sigma}_n$ ratio $\approx 55/45 \approx 1$, is consistent
with independent fits of the zero-field specific
heat~\cite{Europhys} and penetration depth~\cite{Carrington}, and
with band-structure calculations~\cite{Liu,Golubov,Choi}.
Furthermore, $\gamma^{\pi}(H)$ appears to follow quantitatively
numerical simulations~\cite{Nakai,Morten}. This shows that a model
consisting of two superfluids of nearly equal weight, having
different gaps and different anisotropy, gives a consistent
description of various properties of MgB$_2$.

The recent measurements of  the thermal conductivity of single
crystals by Sologubenko {\it et al.}~\cite{Sologubenko} were
interpreted in a similar way. At low fields, the heat conductivity
increases sharply, for all field directions. As discussed for
$\gamma(H)$, this is attributed to the contribution of normal 3D
$\pi$-band carriers in the anomalously large vortex cores. At
higher field, this effect saturates and the anisotropy rises, due
to the increasing contribution of the quasi 2D $\sigma$-band. A
similar approach may explain not only why the anisotropy obtained
from magnetic torque experiment changes with $H$ and $T$, but also
why a single-anisotropy model does not fit the angular dependence
of \Hc~\cite{Angst,Eltsev}. We emphasize, however, that this view
is simplified and phenomenological. A more elaborate description
for the interaction between the two bands was
proposed~\cite{Kogan,Golubovlambda} for the anisotropy of
penetration depth and its temperature dependence (which can be
understood as a consequence of the rapid thermal depletion of the
isotropic $\pi$-band condensate).

The phenomenological two-band model of Ref.~\cite{Europhys} also
uses two independent superfluids to explain the zero-field $C$ of
MgB$_2$. Inter-band coupling was introduced by considering that
the ``virtual'' $T_c$ of the $\pi$-band
($\Delta_{\pi}/1.76\text{k}_{\text{B}} \sim 13$~K) is brought up
to that of the $\sigma$-band; the peculiar features in $C$ then
originate from this superconductivity ``above $T_c$''. Here a
somewhat similar situation arises: a ``virtual'' upper critical
field can be defined for the $\pi$-band, $H_{c2}^{\pi} \sim
(\Phi_0 \Delta_{\pi}^2)/(\hbar^2 v_F^2)$~\cite{ShulgaHcd};
superconductivity persists up to \Hcc or \Hcab, depending on the
field orientation, but above $H_{c2}^{\pi}$ the overlap of huge
vortex cores drives the majority of the $\pi$-band electrons
normal (superconductivity ``above \Hc''). Three characteristic
fields are thus needed to describe the physics of MgB$_2$: the two
upper-critical fields \Hcc and \Hcab associated with the
anisotropic $\sigma$-band, and a cross-over field $H_{c2}^{\pi}$
associated with the $\pi$-band. We estimate the latter one by
linearly extrapolating the low field $\gamma(H)$ to the partial
normal-state $\gamma$ for this band ($\sim 0.55\gamma_n$), by
analogy with classic superconductors. The inset of
Fig.~\ref{Figgamma} shows this construction: we find
$\mu_0H_{c2}^{\pi} \approx$~0.3--0.4~T. These fields determine the
values of the three coherence lengths of the system: for the
anisotropic $\sigma$-band $\xi_{\sigma}^{(ab)}\sim 10$~nm and
$\xi_{\sigma}^{(c)}\sim 2$~nm; for the isotropic $\pi$-band
$\xi_{\pi} \sim 30$~nm. The latter value compares favorably with a
direct measurement by STM giving 50~nm for the vortex-core
diameter in the $\pi$-band~\cite{Morten}. All {\it three} fields
are needed to give a phenomenological description of the
superconducting properties of MgB$_2$. Note that the ratio between
$H_{c2}^{(ab)}$ and $H_{c2}^{(c)}$ is driven by $\Gamma_{\sigma}$,
i.e. the anisotropy of the effective masses of the $\sigma$-band,
whereas the ratio between $H_{c2}^{(c)} \sim (\Phi_0
\Delta_{\sigma}^2)/(\hbar^2 v_F^2)$ and $H_{c2}^{\pi}$ is given by
$\sim \Delta_{\sigma}^2/ \Delta_{\pi}^2$, since the Fermi
velocities are comparable on both bands. The experimental ratio
$H_{c2}^{(c)}/H_{c2}^{\pi}\sim 10$ reported here agrees with
various estimations of $\Delta_{\sigma} /\Delta_{\pi}\sim
3$--3.5~\cite{Liu,Europhys,Carrington,Chen,Quilty,Tsuda,Giubileo,%
Szabo,Laube,Golubov,Choi}.

Stimulating discussions with M.~R.~Eskildsen, T.~Dahm, C.~Berthod,
and C.~Marcenat are gratefully acknowledged. This work was
supported by the Swiss National Science Foundation through the
National Centre of Competence in Research ``Materials with Novel
Electronic Properties-MaNEP'' and the New Energy and Industrial
Technology Development Organization (NEDO).

\end{document}